\documentclass[epsf,showpacs,preprintnumbers,amsmath,amssymb]{revtex4}

\newcommand{\simgt}{\lower.5ex\hbox{$\; \buildrel > \over \sim \;$}}
\newcommand{\simlt}{\lower.5ex\hbox{$\; \buildrel < \over \sim \;$}}

\usepackage{graphicx}
\usepackage{dcolumn}
\usepackage{bm}

\begin{document}


\title{Path Integral Formulation for Wave Effect in Multi-lens System}

\author{Kazuhiro Yamamoto}
\affiliation{
Department of Physical Science, Hiroshima University,
Higashi-Hiroshima 739-8526,~Japan
}


\begin{abstract}
A formula to investigate wave effect in multi-lens system is presented
on the basis of path integral formalism by generalizing the work
by Nakamura and Deguchi (1999). Wave effect of a system with two lenses is 
investigated in an analytic way as a simple application to demonstrate 
usefulness of the formula and variety of wave effect in multi-lens 
system. 
\end{abstract}
\maketitle


\def\M{{M}}
\section{Introduction}
Recently several authors have investigated wave effect
in gravitational lensing phenomenon, motivated by the possibility
that the wave effect might be detected in future gravitational 
wave observatories \cite{TN,Seto,YT,PINQ,BHN,ND,TTN}.
The wave effect can be investigated by solving wave equation, in principle.
However, it is difficult to solve the wave equation analytically 
for general configuration of lenses, excepting the simple case of a single 
Schwarzschild lens (e.g., \cite{DeguchiA,DeguchiB}, see also \cite{SEF} 
and references therein, cf.~\cite{YT}). 
Indeed, investigation of the wave effect so far is restricted 
to special case of single lens model with the spherically 
symmetry, i.e., the Schwarzschild lens model and the singular 
isothermal sphere lens model (see \cite{TN}).

Nakamura and Deguchi developed an elegant formalism for  
gravitational lens using the path integral approach \cite{ND}.
The primary purpose of the present paper is to derive a 
useful formula to investigate the wave effect in general multi-lens 
system, by generalizing the formalism by Nakamura and Deguchi. 
We also apply it to a system with two Schwarzschild lenses to 
demonstrate usefulness of the formula.
This paper is organized as follows:
In \S. 2, we present a generalized formula for multi-lens system.
In \S. 3, an application of the formula to a simple
configuration with two lenses is considered. 
\S. 4 is devoted to summary and conclusion.
We use the convention $c=1$.

\section{Generalized Formulation}
\def\dls{{D_{\rm LS}}}
\def\dos{{D_{\rm OS}}}
\def\bftheta{{\Theta}}
\def\calD{{\cal D}}

We start by reviewing the path integral formalism for gravitational lens 
phenomenon \cite{ND}.
We consider the Newtonian spacetime with the metric
\begin{eqnarray}
  ds^2=-(1+2U(r,{\theta},\varphi))dt^2+(1-2U(r,\theta,\varphi))(dr^2+r^2d\theta^2+r^2\sin^2\theta d\varphi^2),
\end{eqnarray}
where $U(r,\theta,\varphi)$ is the Newtonian potential.
Propagation of massless field $\phi$ is described 
by the wave equation
\begin{equation}
  {\partial \over \partial x^\mu}\left(
  \sqrt{-g} g^{\mu\nu} {\partial \over \partial x^\nu}
  \right)\phi({\bf r},t)=0,
\label{WE}
\end{equation}
where $g$ is the determinant of the metric and ${\bf r}$ represents the
spatial coordinates $(r,\theta,\varphi)$.
We consider a monochromatic wave from a point source with 
the wave number $k$. We set
\begin{eqnarray} 
  \phi({\bf r},t) ={A\over r} F({\bf r}) e^{-ik(t-r)},
\end{eqnarray}
where $A$ is a constant, then Eq.~(\ref{WE}) yields
\begin{eqnarray}
  {\partial ^2 F \over \partial r^2}+2ik {\partial F\over \partial r}
  +{1\over r^2} {1\over \sin\theta} {\partial \over \partial \theta}
  \left(\sin\theta {\partial F \over \partial \theta}\right)
  +{1\over r^2} {1\over \sin^2\theta} {\partial^2 F \over \partial \varphi^2}
  -4k^2 U(r,\theta,\varphi) F=0.
\label{WEB}
\end{eqnarray}
Assuming that the first term is negligible compared with the second term
and $\theta\ll1$, Eq.~(\ref{WEB}) reduces to 
\begin{eqnarray}
  i{\partial F\over \partial r}=
  -{1\over 2kr^2} \left[{1\over \theta} {\partial \over \partial \theta}
  \left(\theta {\partial F \over \partial \theta}\right)
  +{1\over \theta^2} {\partial^2 F \over \partial \varphi^2}\right]
  +2k U(r,\theta,\varphi) F.
\label{WEC}
\end{eqnarray}
Due to analogy of Eq.~(\ref{WEC}) with the Schr$\ddot{\rm o}$dinger equation,
using the path integral formulation of quantum mechanics, 
the solution can be written as follows,
\begin{eqnarray}
  F(r,\theta,\varphi)=\int \calD\bftheta \exp \left[ik\int_0^{r}dr \left(
  {1\over 2} r^2 \dot\bftheta^2(r) -2U(r,\bftheta(r))\right)\right],
\label{WED}
\end{eqnarray}
where the dot denotes the differentiation with respect to $r$ 
and $\Theta$ is used to represent the variables $(\theta,\varphi)$, 
which are related by $\Theta=(\theta\cos\varphi,\theta\sin\varphi)$. 
The expression (\ref{WED}) means the sum of all possible path 
$\Theta(r)$, fixing the initial point (source's position) and the 
final point (observer's position).  

Let us consider multi-lens system, including $n$ lenses as shown in Fig.~1,
in which the source is located 
at the origin of coordinate and an observer is located at $(r,\Theta)
=(r_N,\Theta_N)$. The radius between the source and the observer
is discretized by $N$ segments, and each discrete radius from the 
source is labeled by $r_j$ with $j$ from $1$ to $N$. Separation neighboring 
two discrete radii is $\epsilon$. We assume that the $n$ lenses are 
located at the radius $r=r_{l_m}$ with $m$ from $1$ to $n$, and that
the thin lens approximation is valid for the lenses. In this case the 
explicit expression of Eq.~(\ref{WED}) is written as Eq.~(\ref{FA})
in Appendix (see also \cite{ND}). After some computation, the path 
integral expression reduces to (see Appendix for details)
\begin{eqnarray}
 F(r_N,\Theta_N)&=& {k\over 2\pi i}{r_{l_1}r_{l_2}\over r_{l_1,l_2}}
  \int d^2\Theta_{l_1}  \exp\left[ik\left({r_{l_1}r_{l_2}\over 2r_{l_1,l_2}}
  |\Theta_{l_1}-\Theta_{l_2}|^2-\psi(\Theta_{l_1})\right)\right]
\nonumber
\\
  &\times& {{k\over 2\pi i}}{r_{l_2}r_{l_3}\over r_{l_2,l_3}}
  \int d^2\Theta_{l_2}  \exp\left[ik\left({r_{l_2}r_{l_3}\over 2r_{l_2,l_3}}
  |\Theta_{l_2}-\Theta_{l_3}|^2-\psi(\Theta_{l_2})\right)\right]
\nonumber
\\
  && \hspace{5cm}\cdot
\nonumber
\\
  && \hspace{5cm}\cdot
\nonumber
\\
  && \hspace{5cm}\cdot
\nonumber
\\
  &\times& {{k\over 2\pi i}}{r_{l_{n}}r_N \over r_{l_{n},N}}
  \int d^2\Theta_{l_{n}}  \exp\left[ik\left({r_{l_{n}}r_N\over 2r_{l_{n},N}}
  |\Theta_{l_{n}}-\Theta_N|^2-\psi(\Theta_{l_{n}})\right)\right]
\label{Gen}
\end{eqnarray}
with
\begin{eqnarray}
  \psi(\Theta_{l_m})= 2 \int_{r_{l_m}-\delta r}^{r_{l_m}+\delta r} dr U(r,\Theta),
\end{eqnarray}
where $r_{l_{m-1},l_{m}}=r_{l_m}-r_{l_{m-1}}$, $r_{l_{n+1}}=r_N$, and $\Theta_{l_m}$ is the
variable on the $m$th lens plane. 
%
As expected, stationary condition of the phase of the integrant in Eq.~(\ref{Gen})
reproduces the lens equation in the multi-lens system \cite{SEF}
\begin{eqnarray}
  \nabla_{\Theta_{l_{m'}}} \sum_{m=1}^n \left({r_{l_{m}}r_{l_{m+1}}\over 2r_{l_{m},l_{m+1}}} 
  |\Theta_{l_{m}}-\Theta_{l_{m+1}}|^2-\psi(\Theta_{l_{m}})\right)=0,
\end{eqnarray}
for each $m'=1,\cdots,n$, which hints at a way to the geometrical optics limit.
The Gaussian approximation around a stationary solution yields the
result in the geometrical optics limit (see also \cite{ND}).  


\section{Application to a simple configuration}
In this section we consider a simple case with two lenses. 
Fig.~2 shows the configuration: The Schwarzschild lenses 
with mass $M_1$ and $M_2$ are located at the radius $r_1$
and $r_2$, respectively. The model considered here is not
general, because the source and the two lenses are
arranged to be on a straight line. However, this 
simplification allows us to perform integration of 
Eq.~(\ref{Gen}) analytically. We start by rewriting 
Eq.~(\ref{Gen}) 
\begin{eqnarray}
 F&=& {k\over 2\pi i}{r_{1}r_{2}\over r_{1,2}}
  \int d^2\Theta_{1}  \exp\left[ik\left({r_{1}r_{2}\over 2r_{1,2}}
  |\Theta_{1}-\Theta_{2}|^2-\psi(\Theta_{1})\right)\right]
\nonumber
\\
  &\times& {{k\over 2\pi i}}{r_{2}r_{3}\over r_{2,3}}
  \int d^2\Theta_{2}  \exp\left[ik\left({r_{2}r_{3}\over 2r_{2,3}}
  |\Theta_{2}-\Theta_{3}|^2-\psi(\Theta_{2})\right)\right]
\label{Gen2}
\end{eqnarray}
with
\begin{eqnarray}
  &&\psi(\Theta_{j})= {4 GM_j} \ln\left(|\Theta_{j}|\right), 
\end{eqnarray}
for $j=1,2$, where $r_{i,j}=r_j-r_i$ and 
the possition of an observer is $(r_3,\Theta_3)$. From Eq.~(\ref{Gen2})
we have 
\begin{eqnarray}
 F&=& {k\over i}{r_{1}r_{2}\over r_{1,2}}
  \int_0^\infty d\theta_{1}\theta_{1}^{1-4ikGM_1}  
  \exp\left[ik{r_{1}r_{2}\over 2r_{1,2}}
  \left(\theta_{1}^2+\theta_{2}^2\right)\right] 
  J_0\left({kr_1r_2\over r_{1,2}}\theta_1\theta_2\right)
\nonumber
\\
  &\times&{k\over i}{r_{2}r_{3}\over r_{2,3}}
  \int_0^\infty d\theta_{2}\theta_{2}^{1-4ikGM_2}  
  \exp\left[ik{r_{2}r_{3}\over 2r_{2,3}}
  \left(\theta_{2}^2+\theta_{3}^2\right)\right] 
  J_0\left({kr_2r_3\over r_{2,3}}\theta_2\theta_3\right),
\end{eqnarray}
where $|\Theta_j|=\theta_j$ and $J_0(y)$ is the Bessel function of the first kind.
Integration with respect to $\theta_1$ can be performed
(see \cite{SEF})
\begin{eqnarray}
  F&=&e^{i\alpha} e^{\pi k G M_1}\Gamma(1-2ikGM_1)
  {k\over i}{r_{2}r_{3}\over r_{2,3}}
  \int_0^\infty d\theta_{2}\theta_{2}^{1-4ikGM_2}  
\nonumber
\\
  &&\times
  \exp\left[ik\left({r_{1}r_{2}\over 2r_{1,2}}
  +{r_{2}r_{3}\over 2r_{2,3}}\right)\theta_{2}^2\right]
  {}_1F_1\left(1-2ikGM_1,1~;{-ikr_1r_{2}\over 2r_{1,2}}\theta_2^2\right)
  J_0\left({kr_2r_3\over r_{2,3}}\theta_2\theta_3\right),
\nonumber
\\
\end{eqnarray}
where $\alpha$ is a real number which represents a constant phase 
and ${}_1F_1(a,b~;y)$ is the Kummer's function.
With the use of the definition of the Bessel function
\begin{eqnarray}
  J_0(z)=\sum_{L=0}^\infty{(-1)^L\over (L!)^2}\left(
  {z\over 2}\right)^{2L},
\end{eqnarray}
we have (see \cite{Mag})
\begin{eqnarray}
 F&=&e^{i\alpha'}e^{\pi k G (M_1+M_2)}\Gamma(1-2ikGM_1) 
    ~z\sum_{L=0}^\infty{(-i)^L\over (L!)^2} \Gamma(1+L-2ikGM_2)
\nonumber
\\
  &&\times
  (xz)^L
  {}_2F_1\left(1-2ikGM_1,1+L-2ikGM_2,1~;
  1-z\right),
\label{Fgen}
\end{eqnarray}
where  we defined
\begin{eqnarray}
  &&z={r_3(r_2-r_1)\over r_2(r_3-r_1)},
\\
  &&x={kr_2r_3\theta_3^2\over 2(r_3-r_2)},
\end{eqnarray}
$\alpha'$ is a real constant and ${}_2F_1(a,b,c~;y)$ is the Hypergeometric function. 
%
%
In the limit $\theta_3=0~(x=0)$, Eq.~(\ref{Fgen}) reduces to
\begin{eqnarray}
  F&=& e^{i\alpha'}e^{\pi kG(M_1+M_2)} \Gamma(1-2ikGM_1)\Gamma(1-2ikGM_2)
\nonumber
\\
  &&\times z {}_2F_1\left(1-2ikGM_1,1-2ikGM_2,1~;1-z\right),
\end{eqnarray}
and we have
\begin{eqnarray}
  |F|^2&=& {4\pi kGM_1\over 1-e^{-4\pi kGM_1}} {4\pi kGM_2\over 1-e^{-4\pi kGM_2}}
  z^2 \left|{}_2F_1\left(1-2ikGM_1,1-2ikGM_2,1~;1-z\right)\right|^2.
\end{eqnarray}

We consider the coincidence limit that the distance between the two 
lenses becomes zero, i.e., $r_1=r_2$. In this case, from previous
investigation (e.g., \cite{SEF}), $F$ should be
\begin{eqnarray}
 {\cal F}(x)&\equiv&e^{\pi k G (M_1+M_2)}\Gamma(1-2ikGM_1)  
 {}_1F_1\left(1-2ikG(M_1+M_2),1~;-ix\right),
\label{calF}
\end{eqnarray}
excepting a constant phase factor. We compare our result with Eq.~(\ref{calF}). 
First, let us consider the limit $r_1=r_2$, i.e., $z=0$ of Eq.~(\ref{Fgen}). 
Using the mathematical formula
\begin{eqnarray}
  &&\Gamma(1-2ikGM_1)\Gamma(1-2ikGM_2)
  z{}_2F_1\left(1-2ikGM_1,1-2ikGM_2,1~;1-z\right)
\nonumber
\\
 &&\hspace{2cm} = z^{2ikG(M_1+M_2)}{\Gamma(1-2ikG(M_1+M_2))}
\nonumber
\\
 &&\hspace{2cm}
  \times{}_2F_1\left(2ikGM_1,2ikGM_2,-1+2ikG(M_1+M_2)~;z\right)
\nonumber
\\
  &&\hspace{2cm}+z
  {\Gamma(1-2ikGM_1)\Gamma(1-2ikGM_2)\Gamma(-1+2ikG(M_1+M_2))\over
   \Gamma(2ikGM_1)\Gamma(2ikGM_2)}
\nonumber
\\
  &&\hspace{2cm}
  \times{}_2F_1\left(1-2ikGM_1,1-2ikGM_2,2-2ikG(M_1+M_2)~;z\right),
\label{Fgenn}
\end{eqnarray}
in the case $\theta_3=0~(x=0)$, we can easily have 
\begin{eqnarray}
\lim_{z\rightarrow 0}|F|^2=
{4\pi kG(M_1+M_2)\over 1-e^{-4\pi kG(M_1+M_2)}}=|{\cal F}(0)|^2. 
\end{eqnarray}
This is the expected result. Note that $|F|^2$ is regarded as the
magnification factor. 
Fig.~3 plots $R=|F/{\cal F}(0)|$, as a function of $z$, where we 
fixed $kGM_1=kGM_2=1$. Thus the maximum magnification
depends on lens configuration significantly. Fig.~4 plots 
$R=|F/{\cal F}(x)|$ as a function of $x$ when fixing $z=0.5$ 
and $kGM_1=kGM_2=1$, which indicates that interference
pattern also depends on lens configuration \cite{Math}. 
In these figures
we have fixed $kGM_1=kGM_2=1$, but result depends significantly 
on the parameters $kGM_1$, $kGM_2$ and $z$, as shown in Fig.~4, 
which suggests variety of wave effect depending on lens configuration.


\section{Summary and Conclusion}

In the present paper, we have presented a general formula to investigate
wave effect in multi-lens system, which has been derived with the 
use of path integral approach. The formula is expressed in terms 
of integration with respect to variables of lens planes. 
It is difficult to perform the integration in general cases, 
but a system with two Schwarzschild lenses is an example 
for which the integration can be performed in an analytic way. 
The model of considered in the present paper is 
simplified and limited, however, it suggests variety of 
wave effect in multi-lens phenomenon. It is required to 
develop a numerical method to perform integration in general lens 
configuration in future work.

\vspace{1mm}
\begin{acknowledgments} 
This work is supported in part by Grant-in-Aid for
Scientific research of Japanese Ministry of Education, Culture, Sports, 
Science and Technology, No.15740155.
\end{acknowledgments}


\begin{appendix}
\section{Derivation of Eq.~(\ref{Gen})}
In this Appendix, we review derivation of Eq.~(\ref{Gen}) from Eq.~(\ref{WED}). 
We consider the configuration depicted as Fig.~1. The source is located
at the origin of coordinate and the position of an observer is specified
by $(r,\Theta)=(r_N,\Theta_N)$. The space between the source and the
observer is discritized by $N-1$ planes. The radius of each plane is labeled 
by $r_{j}$ for $j=1,\cdots,N-1$. Note that $r_N$ specifies the plane of the observer. 
$\Theta_j$ is (angle) variable on the $j$th plane. We consider the system 
with $n$ lenses, and assume that $m$th lens is located at the radius $r_{l_m}$. 
Assuming the validity of the thin lens approximation, 
we introduce the two dimensional potential 
\begin{eqnarray}
  \psi(\Theta_{l_m})=2\int_{r_l-\delta r}^{r_l+\delta r} dr U(r,\Theta),
\end{eqnarray}
for $m=1,\cdots,n$, respectively. 
In this case, the path integral formula (\ref{WED}) can be written as
\begin{eqnarray}
  F&=&\left[\prod_{j=1}^{N-1}\int {d^2\Theta_{j}\over A_{j}}\right]
    \exp\left[ik\left(\epsilon\sum_{j=1}^{N-1}
    {r_{j}r_{j+1}\over 2}\left|{\Theta_{j+1}-\Theta_{j}\over \epsilon}
    \right|^2-\sum_{m=1}^{n}\psi(\Theta_{l_m})\right)\right],
\label{FA}
\end{eqnarray}
where the normalization is chosen
\begin{eqnarray}
  A_{j}={2\pi i\epsilon\over k r_{j}r_{j+1}},
\end{eqnarray}
so that $F=1$ in the limit $\psi=0$. 
Then, Eq.~(\ref{FA}) is rephrased as
\begin{eqnarray}
   F&=&\left[\prod_{j=1}^{l_1-1}\int {d^2\Theta_{j}\over A_{j}}\right]
    \exp\left[ik\left(\epsilon\sum_{j=1}^{l_1-1}
    {r_{j}r_{j+1}\over 2}\left|{\Theta_{j+1}-\Theta_{j}\over \epsilon}
    \right|^2\right)\right]
\nonumber
\\&\times&
    \left[\prod_{j=l_1}^{l_2-1}\int {d^2\Theta_{j}\over A_{j}}\right]
    \exp\left[ik\left(\epsilon\sum_{j=l_1}^{l_2-1}
    {r_{j}r_{j+1}\over 2}\left|{\Theta_{j+1}-\Theta_{j}\over \epsilon}
    \right|^2-\psi(\Theta_{l_1})\right)\right]
\nonumber
\\
  && \hspace{3cm}\cdot
\nonumber
\\
  && \hspace{3cm}\cdot
\nonumber
\\
  && \hspace{3cm}\cdot
\nonumber
\\
  &\times&
    \left[\prod_{j=l_n}^{N-1}\int {d^2\Theta_{j}\over A_{j}}\right]
    \exp\left[ik\left(\epsilon\sum_{j=l_n}^{N-1}
    {r_j r_{j+1}\over 2}\left|{\Theta_{j+1}-\Theta_{j}\over \epsilon}
    \right|^2-\psi(\Theta_{l_{n}})\right)\right],
\label{ApA}
\end{eqnarray}
where $\epsilon$ is the separation between two neighboring planes. 
With the use of the following equality, which can be proven by the
mathematical induction, 
\begin{eqnarray}
  &&\sum_{j=l_m}^{l_{m+1}-1} r_{j}r_{j+1}\left|
  \Theta_{j+1}-\Theta_j\right|^2
\nonumber
\\  &&\hspace{2cm}=\epsilon{r_{l_m}r_{l_{m+1}}\over r_{l_{m+1}}-r_{l_m}} 
  \left|\Theta_{l_{m+1}}-\Theta_{l_{m}}\right|^2
  +\sum_{j=l_m+1}^{l_{m+1}-1}r_{j}^2{r_{j+1}-r_{l_m}\over 
  r_j-r_{l_m}}
  \left|\Theta_{j}-u_{l_m,j}\right|^2
\end{eqnarray}
with
\begin{eqnarray}
  u_{l_m,j}={r_{l_m}\Theta_{l_m}+(j-l_m)r_{j+1}\Theta_{j+1}\over 
  j(r_{j+1}-r_{l_m})},
\end{eqnarray}
we have
\begin{eqnarray}
   F&=&
    \int {d^2\Theta_{l_1}}
    {k\over 2\pi i}{r_{l_1}r_{l_2}\over (r_{l_2}-r_{l_1})}
    \exp\left[ik\left(
    {r_{l_1}r_{l_2}\over 2(r_{l_2}-r_{l_1})}\left|{\Theta_{l_2}-\Theta_{l_1}}
    \right|^2-\psi(\Theta_{l_1})\right)\right]
\nonumber
\\
&\times&
    \int {d^2\Theta_{l_2}}
    {k\over 2\pi i}{r_{l_2}r_{l_3}\over (r_{l_3}-r_{l_2})}
    \exp\left[ik\left(
    {r_{l_2}r_{l_3}\over 2(r_{l_3}-r_{l_2})}\left|{\Theta_{l_3}-\Theta_{l_2}}
    \right|^2-\psi(\Theta_{l_2})\right)\right]
\nonumber
\\
  && \hspace{3cm}\cdot
\nonumber
\\
  && \hspace{3cm}\cdot
\nonumber
\\
  && \hspace{3cm}\cdot
\nonumber
\\
  &\times&
    \int {d^2\Theta_{l_n}}
    {k\over 2\pi i}{r_{l_n}r_{N}\over (r_{N}-r_{l_n})}
    \exp\left[ik\left(
    {r_{l_{n}}r_{N}\over 2(r_{N}-r_{l_n})}\left|{\Theta_{N}-\Theta_{l_n}}
    \right|^2-\psi(\Theta_{l_n})\right)\right],
\label{ApA}
\end{eqnarray}
which is equivalent to Eq.~(\ref{Gen})
\end{appendix}

\begin{figure*}
\includegraphics{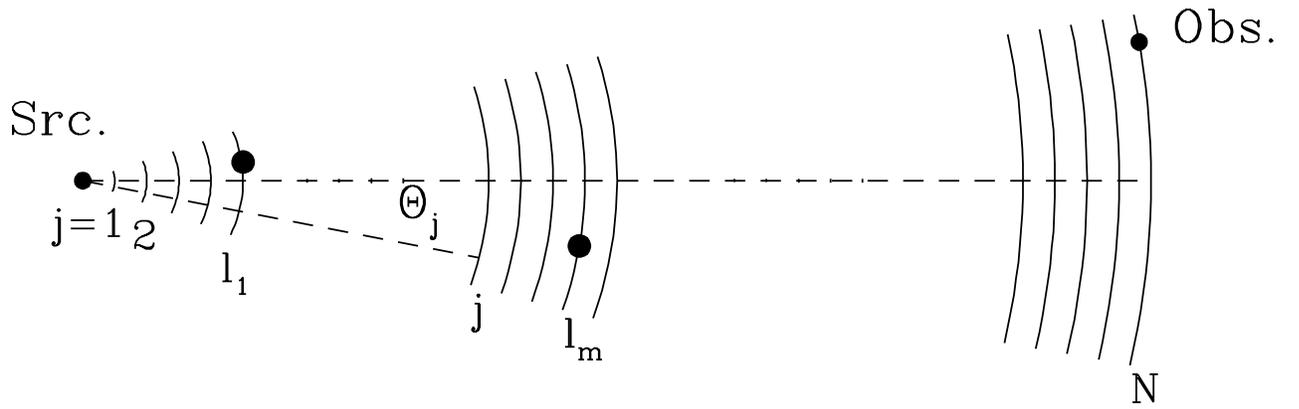}
\caption{\label{fig1} Configuration of multi-lens system 
and coordinates for path integral formula.}
\end{figure*}

\begin{figure*}
\includegraphics{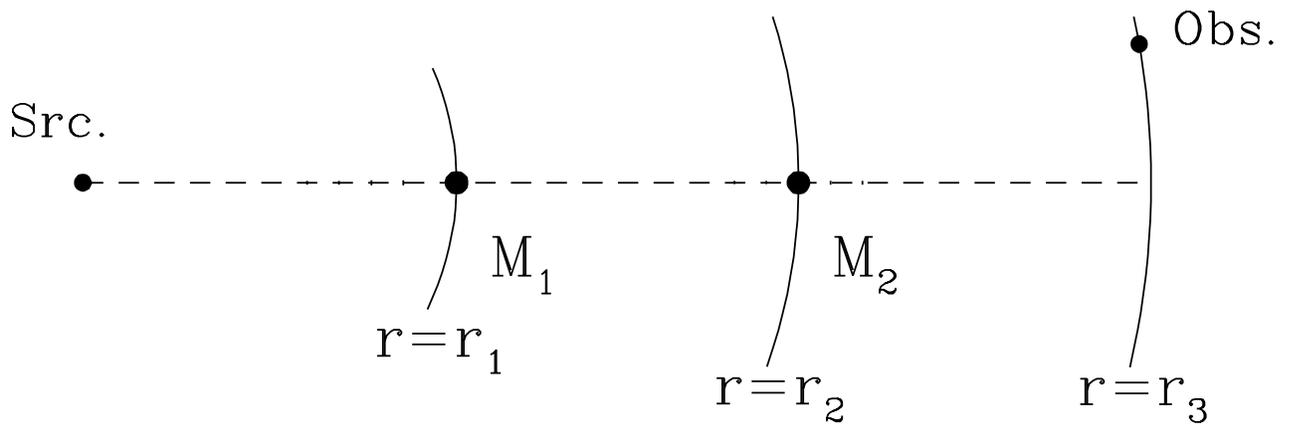}
\caption{\label{fig2} Lensing system considered in section 3.}
\end{figure*}


\begin{figure*}
\includegraphics{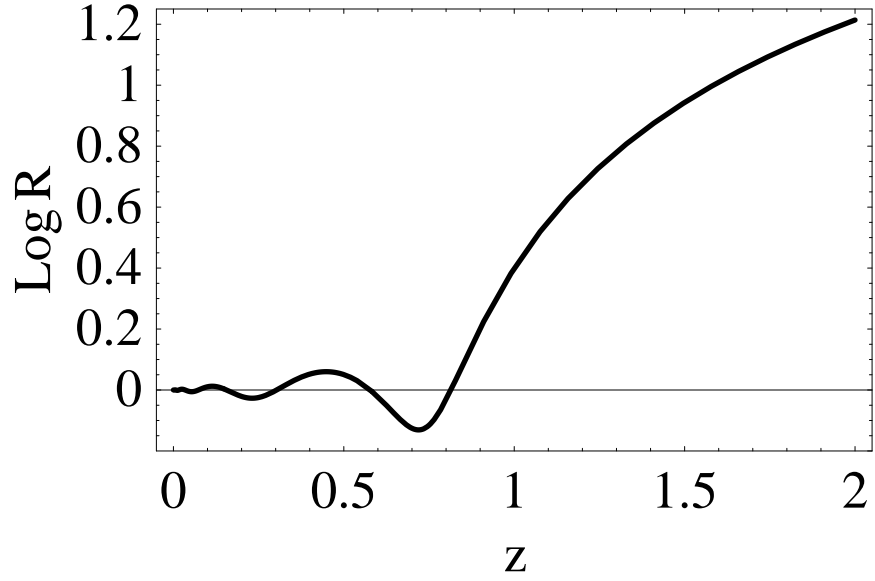}
\caption{\label{fig3} $R=|F/{\cal F}(0)|$ as a function of $z$ 
with fixing $kGM_1=kGM_2=1$.
}
\end{figure*}

\vspace{20cm}

\newpage
\begin{figure}
\includegraphics{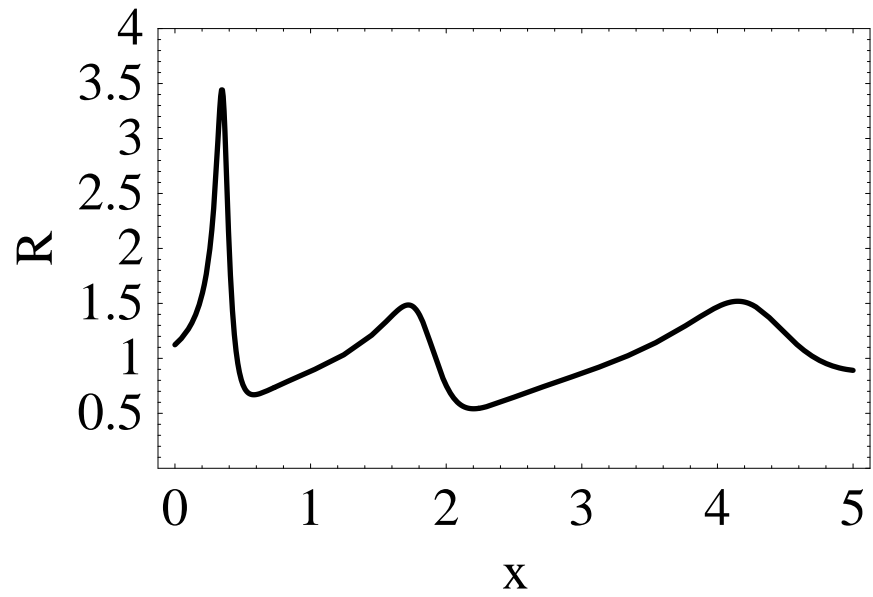}
\caption{\label{fig4} $R=|F/{\cal F}(x)|$ as a function of $x$ with 
fixing $z=0.5$ and $kGM_1=kGM_2=1$.}
\end{figure}


\begin{thebibliography}{99}
\bibitem{Math} The computation was done with the Mathematica.
\end{thebibliography}
\end{document}